\documentclass[12pt]{iopart}
\usepackage{bm}
\usepackage{graphicx}
\usepackage{iopams}
\begin{document}

\title[Confinement induced orbital breathing, fusion, fission and re-ordering]{Confinement induced orbital breathing, fusion, fission and re-ordering in semifilled shell atoms}
\author{ V K Dolmatov}
\address{Department of Physics and Earth Science, University of North Alabama,
Florence, Alabama 35632, USA}
\ead{vkdolmatov@una.edu}
\begin{abstract}
Alternate contraction and drastic expansion, i.e., `breathing' of electronic subshells, the effects of fusion of two subshells into one subshell and its subsequent fission (splitting) into the original subshells, as well as multiple alteration of the order of subshells in confined semifilled shell atoms with a progressively narrowing confinement are theoretically discovered. The confinement is represented by a repulsive penetrable spherical potential of an inner radius $r_{0}$. The effects are exemplified by calculated data for confined semifilled shell atoms from the second, third and fourth rows of Mendeleev's table - Li, N, P and Cr atoms with semifilled ${2\rm s}^{1}$,
${2\rm p}^{3}$, ${3\rm p}^{3}$ and ${3\rm d}^{5}$ subshells, respectively - for the completeness of the study. The underlying physics behind the discovered effects is explained.
\end{abstract}
\pacs{32.80.Fb, 32.30.-r, 31.15.V-}
\submitto{\jpb}
\maketitle
\section{Introduction}

The structure and spectra of atoms, spatially confined by various types of confinements whose sizes are commensurable with an atomic size, have seized the attention of theorists starting since early days
\cite{Michels,Sommerfeld} to now, see review papers \cite{JPC,RPC'04,AQC57,AQC58} as well as some of the latest works on the subject \cite{Dolmatov12,Manson13,Cruz13,Auburn13} and references therein. This is because a confined atom concept provides insights into various aspects of interdisciplinary significance
\cite{JPC,RPC'04,AQC57,AQC58}. The most illustrative examples, perhaps, could be atoms under high pressure, quantum dots and endohedral fullerenes $A$@C$_{n}$.
To date, numerous aspects of the structure and spectra of atoms under various kinds of
confinements - impenetrable, open boundaries, spheroidal, conoidal, Deby, C$_{n}$-like,\textit{etc.} - have been attacked from many different angles by research teams world-wide \cite{JPC,RPC'04,AQC57,AQC58} (and references therein).

The present paper reports on
novel discoveries made, specifically, for the structure of semifilled shell atoms under confinement. The latter is simulated by a repulsive penetrable spherical potential of an adjustable inner radius $r_{0}$.
In such atoms, spectacular confinement induced  effects termed orbital breathing, fusion, fission, and re-ordering with changing $r_{0}$ have been unravelled.
Semifilled shell atoms from the second, third and fourth rows of Mendeleev's
table - Li(...$\rm 2s^{1}$), N(...$\rm 2p^{3}$), P(...$\rm 3p^{3}$) and Cr(...$\rm 3d^{5}$$\rm 4s^{1}$) - are chosen for completeness of the case study. Atomic units (a.u.) are used throughout the paper unless specified otherwise.

When a multielectron atom is placed inside a repulsive spherical potential of an inner radius $r_{0}$, finite height $U_{0}$ and width $\Delta$, its electrons necessarily move in an
effective double-well potential, even if  the original atomic potential was single-welled. A problem of the thus confined multielectron atom brings together two distinct brunches of research in atomic physics.
On the one hand, it features  peculiarities of the structure of electronic subshells in the field of a double-well potential. On the other hand, it studies how multielectron atoms may react to the presence
of confinement. Both of these outstanding topics have attracted much attention of researchers from the early to modern days. Thus, for double-well potential studies - see the original work by Goeppert-Mayer \cite{G-Mayer} and corresponding comprehensive review by Connerade \cite{Connerade_book}. For confined atoms - refer to the above referenced papers \cite{Michels,Sommerfeld,JPC,RPC'04,AQC57,AQC58,Dolmatov12,Manson13,Cruz13,Auburn13} and references therein.

The impetus for the present study stems from a recent work \cite{Dolmatov12} on the ground and excited states of the hydrogen atom confined inside a repulsive spherical potential $V_{\rm c}(r)$ of an inner radius $r_{0}$, finite height $U_{0}$ and width $\Delta$.  Addition of $V_{\rm c}(r)$ to the hydrogen
potential $V_{\rm H}(r)$ resulted in a double-well effective potential $U_{\rm eff} = V_{\rm H}+V_{\rm c}$ of the atom. The binding strengths of the inner and outer wells were controlled by decreasing $r_{0}$.
It was shown that the binding strength of $U_{\rm eff}$ alters in favour of the outer well at $r_{0} \leq r_{\rm c}$, where $r_{\rm c}$ is a certain critical radius. As a result, e.g., the radius of the $1\rm s$ orbital suddenly increased from
only $r_{1\rm s} =0.76$ at $r_{0} = 1.59$  to $r_{1\rm s} \approx 11$ at $r_{0} \leq 1.45$, at certain $U_{0}$ and $\Delta$.
The discovered effect was termed \textit{atomic swelling upon compression}. It was found to alter drastically not only the size and  $\epsilon_{n \ell}$ energy of the H(${n\ell}$) atom but its
oscillator strengths  $f_{{\rm 1s} \rightarrow {n\rm p}}$ as well, either almost extinguishing or drastically increasing  $f_{{\rm 1s} \rightarrow {n\rm p}}$ depending on $r_{0}$.

The above findings naturally make one to wonder how a multielectron atom might behave under the same kind of confinement. The present paper satisfies such curiosity. It is
uncovered, in the performed study, that particularly semifilled shell atoms behave utterly spectacular under said confinement. It is the ultimate aim of this paper to demonstrate
that, as the radius $r_{0}$ of confinement is changing, these atoms may develop such effects as orbital breathing (alternate swelling and collapse of an orbital with changing $r_{0}$), orbital fusion (both energies and radial functions of two orbitals become equal through a certain range of $r_{0}$ values), orbital fission (the two fused together orbitals separate once again through a certain different range of $r_{0}$), as well as orbital re-ordering (the order of orbitals in the atom changes with decreasing $r_{0}$).

\section{Theory in brief}

A convenient way to calculate the structure of a semifilled shell atom is provided by a `spin-polarized' Hartree-Fock (SPHF) approximation \cite{Slater}. SPHF was successfully applied by the author of this paper and his colleagues to said atoms
on numerous occasions to date, see, e.g., \cite{JETP83,Cr93,Mn3s} and references therein. The quintessence of SPHF is as follows. It accounts for the fact that spins of all
electrons in a semifilled subshell of the atom are co-directed, in accordance with Hund's rule, conditionally - all pointing upward ($\uparrow$). This results in splitting of a
closed ${n\ell}^{2(2\ell+1)}$ subshell in the atom into two semifilled subshells of opposite spin orientations, ${n\ell}^{2\ell+1}$$\uparrow$ and ${n\ell}^{2\ell+1}$$\downarrow$, due to
 the presence of
exchange interaction between $nl$$\uparrow$ electrons with only spin-up electrons in the semifilled
subshell(s) of the atom but  absence of such for $nl$$\downarrow$ electrons. Thus, atoms of concern of this paper - Li, N, P and Cr - have  the following SPHF configurations:
 Li(${1\rm s}^{1}$$\uparrow$$ {1\rm s}^{1}$$\downarrow$${2\rm s}^{1}$$\uparrow$),
N(${1\rm s}^{1}$$\uparrow$${1\rm s}^{1}$$\downarrow$${2\rm s}^{1}$$\uparrow$${2\rm s}^{1}$$\downarrow$${2\rm p}^{3}$$\uparrow$),
P(${1\rm s}^{1}$$\uparrow$${1\rm s}^{1}$$\downarrow$${2\rm s}^{1}$$\uparrow$${2\rm s}^{1}$$\downarrow$${2\rm p}^{3}$$\uparrow$${2\rm p}^{3}$$\downarrow$${3\rm s}^{1}$$\uparrow$${3\rm s}^{1}$$\downarrow$${3\rm p}^{3}$$\uparrow$)
and
Cr(...${3\rm p}^{3}$$\uparrow$${3\rm p}^{3}$$\downarrow$${3\rm d}^{5}$$\uparrow$${4\rm s}^{1}$$\uparrow$).
SPHF equations for semifilled shell atoms differ from ordinary HF equations for closed shell atoms by  accounting for the exchange interaction only between electrons with the same spin orientation
($\uparrow$,$\uparrow$ or $\downarrow$,$\downarrow$).

As for a  semifilled shell atom confined inside a spherical potential $U_{\rm c}(r)$, its orbital radial functions $P_{n\ell\uparrow\downarrow}(r)$ and one-electron energies $\epsilon_{n\ell\uparrow\downarrow}$ are the solutions of a system of \textit{modified} radial SPHF equations which account for the presence of $U_{\rm c}(r)$ in addition to the purely atomic SPHF potential:
\begin{equation}
\left.\eqalign{[\hat{H}^{r} + U_{\rm c}(r)]P_{n\ell\uparrow}(r) = \epsilon_{n\ell\uparrow}P_{n\ell\uparrow}(r) \cr
[\hat{H}^{r} + U_{\rm c}(r)]P_{n\ell\downarrow}(r) = \epsilon_{n\ell\downarrow}P_{n\ell\downarrow}(r)}\right\}.
\label{HF}
\end{equation}
Here, $\hat{H}^{r}$ is a radial part of the non-local SPHF Hamiltonian which, naturally, is identical to that for a free atom. In this paper, as was specified above, the confining potential $U_{\rm c}(r)$
is defined by
\begin{eqnarray}
U_{\rm c}(r)=\left\{\matrix {
U_{0}>0, & \mbox{if $r_{0} \le r \le r_{0}+\Delta$} \nonumber \\
0 & \mbox{otherwise,} } \right.
\label{SWP}
\end{eqnarray}
where $U_{0}$, $r_{0}$ and $\Delta$ are its height, inner radius and thickness, respectively.
As in \cite{Dolmatov12}, the height $U_{0}$ and thickness $\Delta$ of the potential will be fixed at $U_{0}=2.5$ and $\Delta=5$, just for the sake of certainty.

\section{Results and discussion}

The author believes that calculated data for a confined phosphorous atom are most illustrative with respect to confinement induced effects emerging in semifilled shell atoms. Therefore,
let us discuss calculated data for confined phosphorous in the first head.

\subsection{The confined P(...${3\rm s}^{1}$$\uparrow$${3\rm s}^{1}$$\downarrow$${3\rm p}^{3}$$\uparrow$)  atom}

SPHF calculated data for the free and confined phosphorous atoms are depicted in figures \ref{fig1} and \ref{fig2} which illustrate all the unravelled effects.

\subsubsection{Orbital breathing.}

\begin{figure}[h]
\center{\includegraphics[width=8cm]{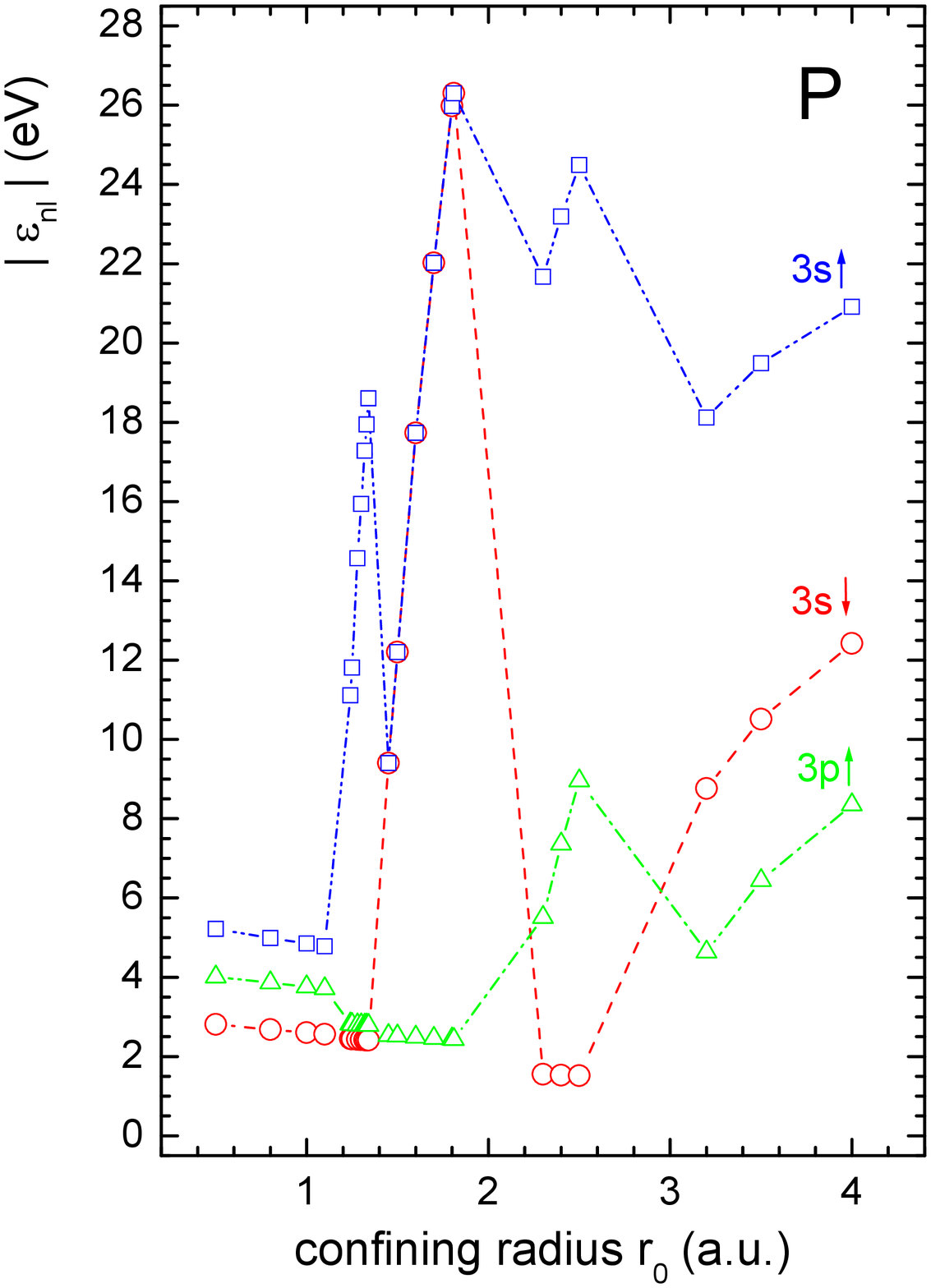}}
\caption{Absolute values of the $\epsilon_{\rm 3s\uparrow}$, $\epsilon_{\rm 3s\downarrow}$ and $\epsilon_{\rm 3p\uparrow}$  energies of
the $\rm 3s$$\uparrow$, $\rm 3s$$\downarrow$ and $\rm 3p$$\uparrow$ orbitals, as marked, of confined phosphorous versus the radius $r_{0}$ of confinement (\ref{SWP}).}
\label{fig1}
\end{figure}

Notice (see figure \ref{fig1}) how $|\epsilon_{3\rm s\downarrow}|$ of the ${3\rm s}$$\downarrow$ orbital first drastically
decreases from $|\epsilon_{3\rm s\downarrow}| = 12.4$ eV at $r_{0}=4$ to only $|\epsilon_{3\rm s\downarrow}| = 1.5$ eV at $r_{0} =2.5$,
then significantly  increases to
$|\epsilon_{3\rm s\downarrow}| = 26.3$ eV at $r_{0} \approx 1.8$, then, once again, drastically  decreases to $|\epsilon_{3\rm s\downarrow}| = 2.4$ eV at $r_{0}=1.34$,
after which it remains about constant at lower values of the confining radius $r_{0}$. The energy variations are accompanied by variations in the size (radius) of the  ${3\rm s}$$\downarrow$
orbital, as clearly follows from figure \ref{fig2} where the radial function  $P_{3\rm s\downarrow}(r)$ is plotted for selected values of $r_{0}$. Indeed, one can see that the ${3\rm s}$$\downarrow$ orbital first dramatically swells (in terms of work \cite{Dolmatov12}) in size at $r_{0} = 2.5$, where the most likely value of its radius is $r_{\rm 3s\downarrow} \approx 13$ versus
of only $r_{\rm 3s\downarrow} \approx 1.6$ in the free atom, then significantly collapses to $r_{\rm 3s\downarrow} \approx 1.2$ at $r_{0} = 1.45$ and considerably swells once again
to $r_{\rm 3s\downarrow} \approx 9.25$ at $r_{0} = 1.34$, after which it remains about the same at smaller values of $r_{0}$.  The emerging variations in $\epsilon_{3\rm s\downarrow}$ and
alternate contractions and expansions of $P_{3\rm s\downarrow}(r)$ with changing confining radius $r_{0}$ are clearly indicative of \textit{orbital `breathing'} - one of the novel effect discovered in the present work.

Also, notice (figure \ref{fig1}) how the $\epsilon_{n\ell\uparrow\downarrow}$ energies of confined phosphorous are alternately increasing and decreasing with changing $r_{0}$ even though
corresponding $n\ell$$\uparrow$$\downarrow$ orbitals remain either spatially swollen or collapsed, at given values of $r_{0}$. This is an interesting observation by itself. For example, one can count seven oscillations in the energy of the  $\rm 3s$$\uparrow$ orbital as $r_{0}$ is changing from $r_{0}=4$ to $r_{0}=0.5$.

\subsubsection{Orbital fusion.} 
\begin{figure}[h]
\center{\includegraphics[width=8cm]{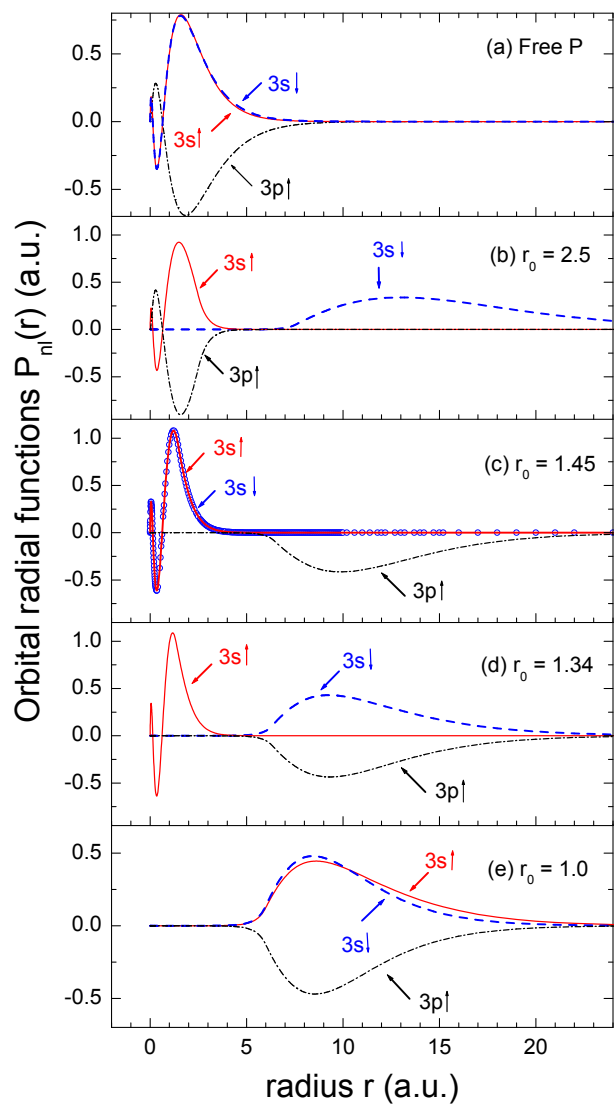}}
\caption{Orbital radial function $P_{\rm 3s\uparrow}(r)$, $P_{\rm 3s\downarrow}(r)$ and $P_{\rm 3p\uparrow}(r)$ of both free and confined phosphorous calculated at selected values of the
 confining radius $r_{0}$, as marked.}
\label{fig2}
\end{figure}

Furthermore, notice (see figure \ref{fig1}) how the one-electron energies $\epsilon_{3\rm s\uparrow}$ and $\epsilon_{3\rm s\downarrow}$,
  which are normally much different from each other (e.g., $\epsilon_{\rm 3s\uparrow} =20.9$ eV whereas $\epsilon_{\rm 3s\uparrow} =12.4$ eV at $r_{0}=4$), become the same, to an excellent approximation,
   in the whole interval of $1.45 \leq r_{0} \leq 1.81$. Not only $\epsilon_{3\rm s\uparrow} =\epsilon_{3\rm s\downarrow}$
in the specified values of the confining radius $r_{0}$, but the radial functions    $P_{3\rm s\uparrow}(r)$ and $P_{3\rm s\downarrow}(r)$ as well. This is
illustrated by  figure \ref{fig2}(c),  where $P_{3\rm s\uparrow}(r)$ and $P_{3\rm s\downarrow}(r)$ are plotted for a selected value of $r_{0}=1.45$. Moreover, calculated data
(not presented in the paper) showed that the same is true for  all other spin-up and spin-down orbitals $n\ell$$\uparrow$ and $n\ell$$\downarrow$ as well, at given $1.45 \leq r_{0} \leq 1.81$.
The emerging equivalence of one-electron energies ($\epsilon_{n\ell\uparrow} = \epsilon_{n\ell\downarrow}$)  and radial functions ($P_{n\ell\uparrow} = P_{n\ell\downarrow}$) of  $n\ell$$\uparrow$ and $n\ell$$\downarrow$ orbitals in a certain range of values of $r_{0}$ is indicative of \textit{orbital `fusion'}, where each of two individual entities - $n\ell^{2\ell+1}$$\uparrow$ and $n\ell^{2\ell+1}$$\downarrow$ subshells - turn into
a single entity - a $n\ell^{2(2\ell+1)}$ subshell.  Correspondingly, the electron configuration of the phosphorous atom with mutually fused spin-up and spin-down orbitals becomes
P(${1\rm s}^{2}{2\rm s}^{2}{2\rm p}^{6}{3\rm s}^{2}{3\rm p}^{3}$), as in a standard HF theory. Orbital fusion is precisely another of the novel effects discovered in the present work.

The physics behind of orbital fusion becomes clear when one explores figure \ref{fig2}(c). One can see that the effect takes place when only the outermost semifilled ${3\rm p}^{3}$$\uparrow$
 subshell of the atom swells into the outer well of the effective potential as a result of changing $r_{0}$. When this happens, the most likely value of the radius of the ${3\rm p}^{3}$$\uparrow$
 orbital increases from its ordinary value of $r_{3\rm p} \approx 1.5$ in the inner well to $r_{3\rm p} \approx 10$ in the outer well, whereas, in contrast, the radii of $\rm 3s$$\uparrow$ and
 $\rm 3s$$\downarrow$ orbitals both shrink to $r_{\rm 3s\uparrow\downarrow} \approx 1.2$. Because of huge differences between the radii of the swollen ${3\rm p}$$\uparrow$
 and all other orbitals of the atom,
exchange interaction of inner electrons with the outermost ${3\rm p}$$\uparrow$ electrons vanishes, to an excellent approximation. This wipes off any differences between the energies and wavefunctions
of inner $n\ell$$\uparrow$ and $n\ell$$\downarrow$ electrons with the same $n\ell$, for an obvious reason. Thus, $n\ell$$\uparrow$$\downarrow$ orbital fusion takes place in the confined atom, at specific values of $r_{0}$.

\subsubsection{Orbital fission.}

As  $r_{0}$ keeps decreasing, the binding strength of the inner well lessens whereas that of the outer well increases. Correspondingly, the latter can support more and more orbitals, with decreasing $r_{0}$. Thus, e.g.,
see figure \ref{fig2}(d), as $r_{0}$ is decreased to $r_{0}=1.34$, the outer well is capable of binding, and does bind, two orbitals - the $\rm 3s$$\downarrow$ and $\rm 3p$$\uparrow$ orbitals. It is, however,
 still not powerful enough to support the third orbital,
 $\rm 3s$$\uparrow$. Therefore,  the $\rm 3s$$\uparrow$ and $\rm 3s$$\downarrow$ orbitals, which were fused together in the range of
 $1.45 \leq r_{0} \leq 1.81$, become  split  once again, both in terms of their radii and energies
($|\epsilon_{3\rm s\uparrow}| \approx 19$ eV versus $|\epsilon_{3\rm s\downarrow}| \approx 2$ eV), at $r_{0}=1.34$.  This is indicative of confinement induced \textit{orbital fission} which is another novel effect
 uncovered in the present work. It is important to underline that confinement induced orbital fission is qualitatively different from the cause of splitting of a ${n\ell}^{2(2\ell+1)}$ subshell into
 spin-up ${n\ell}^{(2\ell+1)}$$\uparrow$ and spin-down ${n\ell}^{(2\ell+1)}$$\downarrow$ subshells suggested by the SPHF theory alone. Indeed, confinement induced orbital fission is exclusively due to orbital swelling
 of the $\rm 3s$$\downarrow$ orbital but shrinking of the $\rm 3s$$\uparrow$ orbital, whereas, in the SPHF theory, the splitting is solely due to differences in exchange interaction of spin-up and spin-down electrons with
 only spin-up electrons in an unpaired semifilled subshell of the atom. Confinement induced orbital fission is another novel effect uncovered in the present work.

\subsubsection{Three-orbital swelling.}

 As $r_{0}$ is decreased to a yet smaller value, $r_{0} \leq 1.1$, even three-orbital swelling occurs, where the three ${3\rm p}^{3}$$\uparrow$, $\rm 3s$$\downarrow$ and $\rm 3p$$\uparrow$ orbitals all swell from the inner well into
 the outer well of the effective potential, see figure \ref{fig2}(e). The unravelled three-orbital swelling effect in a double-well potential is an interesting occurrence which was not
 ever noted in previous works on atomic structures in double-well potentials, to the best of the author's knowledge. Note, in the discussed case of three-orbital swelling, the $\rm 3s$$\uparrow$ and $\rm 3s$$\downarrow$ orbitals are split, both in terms of their radial functions (figure \ref{fig2}(e)) and energies:  $|\epsilon_{3s\uparrow}| \approx 5$ eV versus $|\epsilon_{3s\downarrow}| \approx 3$  eV, see figure \ref{fig1}. This time, the splitting
is in full accord with the SPHF theory - it is due to the presence/absence of exchange interaction between the $\rm 3s$$\uparrow$/$\rm 3s$$\downarrow$ and ${\rm 3p}^{3}$$\uparrow$ electrons in the atom
which is now not a negligible effect, because the radii of these three orbitals are about the same.

\subsubsection{Orbital re-ordering}

Finally, notice (see figure \ref{fig1}) how the order of  the $\rm 3s$$\uparrow$, $\rm 3s$$\downarrow$ and ${\rm 3p}^{3}$$\uparrow$ subshells (orbitals) relative to their energies $\epsilon_{3\rm s\uparrow}$, $\epsilon_{3\rm s\downarrow}$ and $\epsilon_{3\rm p\uparrow}$ alters with changing $r_{0}$ from ($3\rm s$$\uparrow$, $3\rm s$$\downarrow$, ${3\rm p}^{3}$$\uparrow$)  at $r_{0}= 4$ (the same as in the free atom)
to ($3\rm s$$\uparrow$, ${3\rm p}^{3}$$\uparrow$, $3\rm s$$\downarrow$) at $r_{0}=2.5$, to
(${3\rm s}^{2}$, ${3\rm p}^{3}$$\uparrow$) at $r_{0}=1.5$ and back to ($3\rm s$$\uparrow$,  ${3\rm p}^{3}$$\uparrow$, $3\rm s$$\downarrow$) at $r_{0} < 1.34$. Such confinement induced multiple re-ordering of orbitals in the atomic structure is one more important novel effect discovered in the present work. The knowledge of shell ordering is fundamental
to chemistry, because it determines the electronic properties of atoms, such, for example, as valence.

\subsection{The N(...${2\rm s}^{1}$$\uparrow$${2\rm s}^{1}$$\downarrow$${2\rm p}^{3}$$\uparrow$)  atom}

\begin{figure}[h]
\center{\includegraphics[width=8cm]{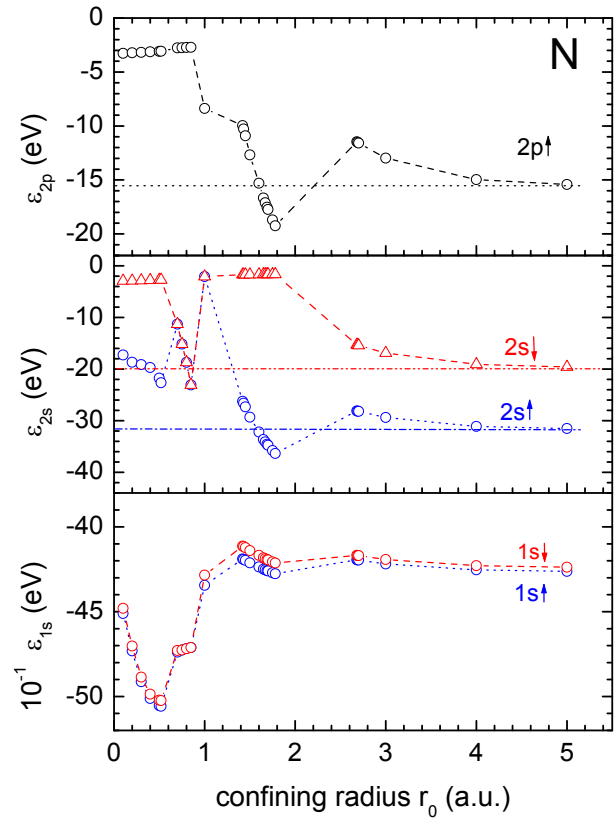}}
\caption{The confined nitrogen $\epsilon_{\rm 1s\uparrow}$,  $\epsilon_{\rm 1s\downarrow}$, $\epsilon_{\rm 2s\uparrow}$, $\epsilon_{\rm 2s\downarrow}$ and $\epsilon_{\rm 2p\uparrow}$  energies of
corresponding orbitals levels, as marked, versus the confining radius $r_{0}$. Horizontal dashed, dotted-dashed and dashed-dotted-dashed lines represent the $\rm 2s$$\uparrow$, $\rm 2s$$\downarrow$
and $\rm 2p$$\uparrow$ energy levels of free nitrogen.}
\label{fig3}
\end{figure}

\begin{figure}[h]
\center{\includegraphics[width=8cm]{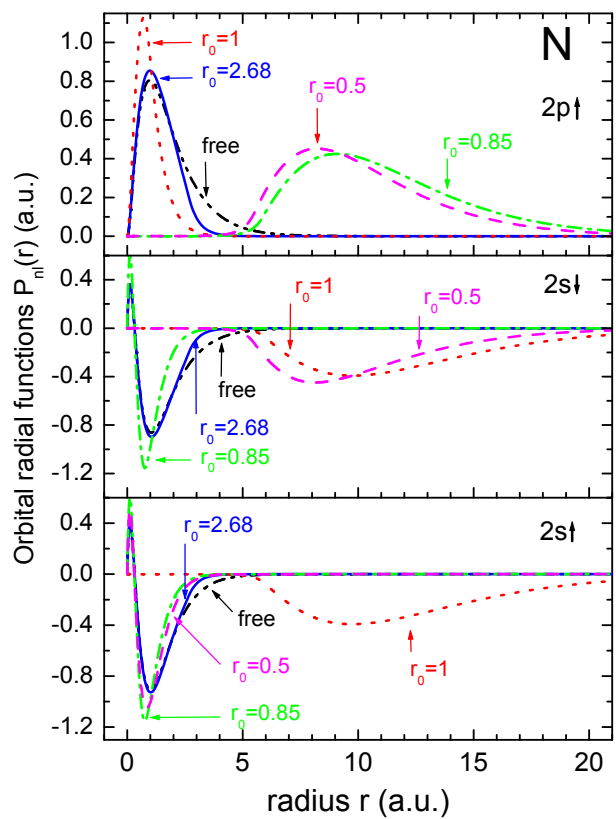}}
\caption{Orbital radial function $P_{\rm 2s\uparrow}(r)$, $P_{\rm 2s\downarrow}(r)$ and $P_{\rm 2p\uparrow}(r)$ of both free and confined nitrogen calculated at selected values of the
 confining radius $r_{0}$, as marked.}
\label{fig4}
\end{figure}

SPHF calculated data for one-electron energies $\epsilon_{n\ell\uparrow\downarrow}$ and orbital radial function $P_{n\ell\uparrow\downarrow}(r)$ of free and confined nitrogen versus the confining radius are depicted in figures
\ref{fig3} and \ref{fig4}. They reveal similar effects of orbital   breathing, fusion, fission and re-ordering as in the above discussed confined phosphorous atom. Notice (figure \ref{fig3}),
how the $2\rm s$$\uparrow$ and  $2\rm s$$\downarrow$ orbitals fuse together in the range of $0.7 \leq r_{0} \leq 1$ but separate
at  $r_{0} < 0.7$, how the subshell ordering (${2\rm s}$$\uparrow$, ${2\rm s}$$\downarrow$, ${2\rm p}^{3}$$\uparrow$) at $r \ge 2.68$ changes to
(${2\rm s}$$\uparrow$, ${2\rm p}^{3}$$\uparrow$, ${2\rm s}$$\downarrow$) at $1.42 \leq r_{0} \leq 1.78$, to  (${2\rm p}^{3}$$\uparrow$, ${2\rm s}^{2}$) at $r_{0}=1$ ($2\rm s$$\uparrow$$\downarrow$ spin-orbitals fuse together and become outermost),
to (${2\rm s}^{2}$, ${2\rm p}^{3}$$\uparrow$) at $0.7 \leq r_{0} \leq 0.85$ ($2\rm s$$\uparrow$$\downarrow$ spin-orbitals remain fused together) and to ($\rm 2s$$\uparrow$, ${\rm 2p}^{3}$$\uparrow$, $\rm 2s$$\downarrow$) at $r_{0} \leq 0.52$. Also notice
orbital fusion of the $\rm 1s$$\uparrow$ and  $\rm 1s$$\downarrow$ orbitals at $0.5 \leq r_{0} \leq 0.85$. Furthermore, notice (figure \ref{fig4}) orbital breathing of the $\rm 2s$$\downarrow$ orbital as $r_{0}$ changes from
$r_{0} =2.68$ to $r_{0}=1$ (the orbital swells in size), then to $r_{0} =0.85$ (the orbital collapses) and to $r_{0}=0.5$ (the orbital swells in size). Orbital breathing is also inherent to the $\rm 2s$$\uparrow$ orbital of the
atom.

\subsection{The Li(${1\rm s}^{1}$$\uparrow$${1\rm s}^{1}$$\downarrow$${2\rm s}^{1}$$\uparrow$) atom}

\begin{figure}[h]
\center{\includegraphics[width=8cm]{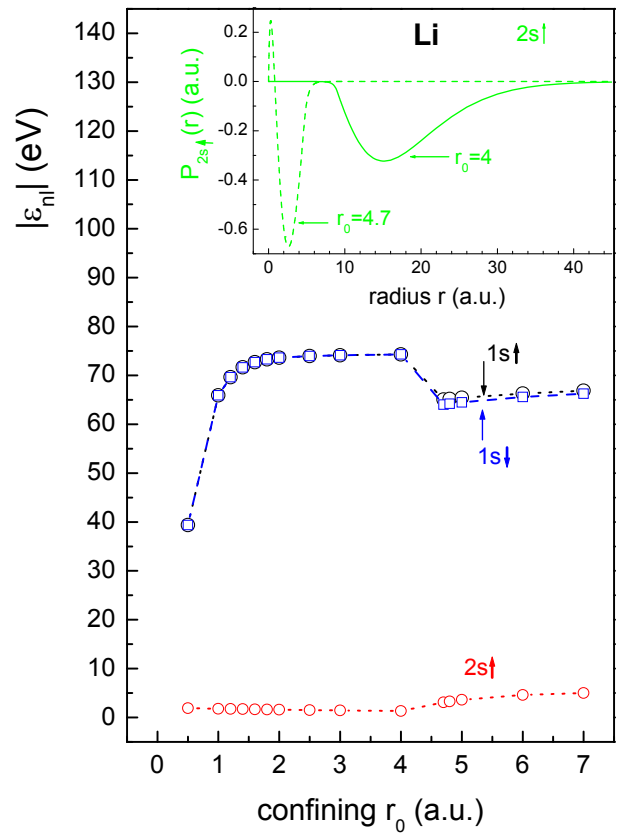}}
\caption{Main panel: $|\epsilon_{\rm 1s\uparrow}|$,  $|\epsilon_{\rm 1s\downarrow}|$ and $|\epsilon_{\rm 2s\uparrow}|$ of confined Li, as marked,
versus the confining radius $r_{0}$. Inset: the orbital radial function $P_{\rm 2s\uparrow}(r)$ of confined Li calculated at $r_{0}= 4.0$ and $4.7$, as marked.}
\label{fig5}
\end{figure}

For the lithium atom, SPHF calculated data (see figure \ref{fig5}) reveal orbital swelling only of  the outermost ${2\rm s}$$\uparrow$ orbital, at $r_{0} \leq 4$. The implication is that the outer well of the confined Li atomic effective potential
is not strong enough to make a ${1\rm s}$$\uparrow$$\downarrow$ electron, or essentially \textit{all} electrons of the atom, to move far away from the nucleus into the outer well of the potential and remain bound.
At $r_{0} \leq 4$, when the $\rm 2s$$\uparrow$ orbital swells into the outer well of the potential,
exchange interaction between the innermost ${1\rm s}$$\uparrow$ electron and the very far distanced outermost  ${2\rm s}$$\uparrow$ electron (both have the same spin orientation) vanishes. This results in orbital fusion of the ${1\rm s}$$\uparrow$ and ${1\rm s}$$\downarrow$ orbitals, as is clearly demonstrated by figure \ref{fig5}. No orbital breathing, fission, or re-ordering was found
to occur in the atom, though.

\subsection{The Cr(...${3\rm p}^{3}$$\uparrow$${3\rm p}^{3}$$\downarrow$${3\rm d}^{5}$$\uparrow$${4\rm s}^{1}$$\uparrow$) atom}

\begin{figure}[h]
\center{\includegraphics[width=8cm]{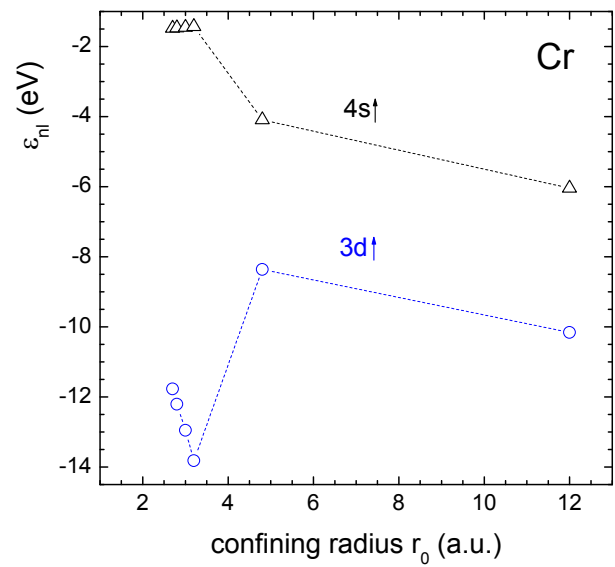}}
\caption{The $\epsilon_{\rm 3d\uparrow}$ and $\epsilon_{\rm 4s\uparrow}$ energies of confined Cr
versus the confining radius $r_{0}$.}
\label{fig6}
\end{figure}

\begin{figure}[h]
\center{\includegraphics[width=8cm]{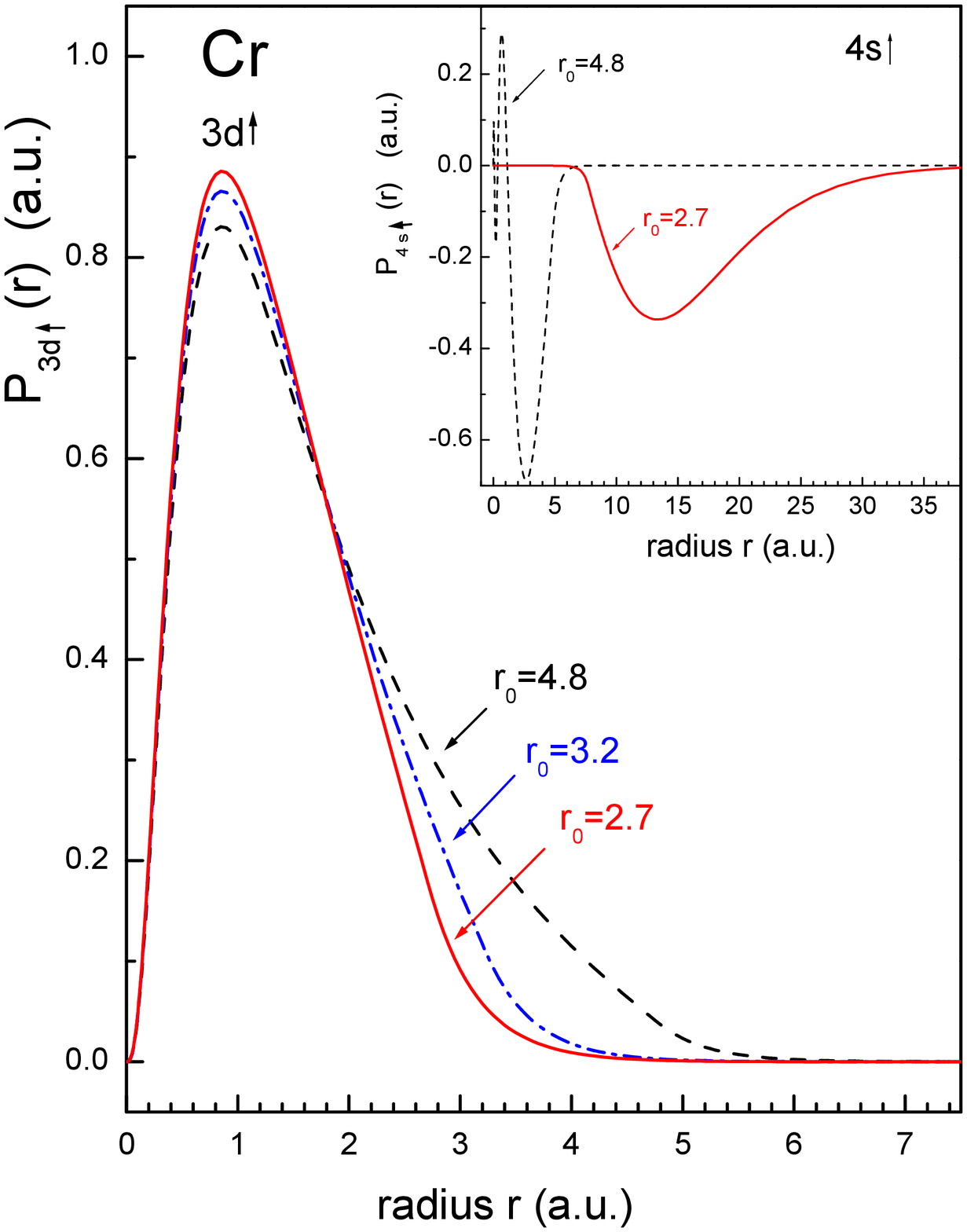}}
\caption{The orbital radial functions $P_{\rm 3d\uparrow}(r)$ (main panel) and $P_{\rm 4s\uparrow}(r)$ (inset) of confined Cr calculated at selected values of the confining radius $r_{0}$, as marked.}
\label{fig7}
\end{figure}

Calculated data for the $\epsilon_{3\rm d\uparrow}$ and $\epsilon_{4\rm s\uparrow}$ energies of confined Cr versus $r_{0}$ are depicted in figures \ref{fig6} and \ref{fig7}.

Notice how $\epsilon_{4\rm s\uparrow}$ and $\epsilon_{3\rm d\uparrow}$, which
behaved identically until $r_{0} \approx 4.8$, start suddenly behaving oppositely at smaller values of $r_{0}$. Indeed, at $r_{0} = 3.2$, $|\epsilon_{4\rm s\uparrow}|$ abruptly decreases from $|\epsilon_{4\rm s\uparrow}| \approx  4$ to $|\epsilon_{4\rm s\uparrow}| \approx 1.4$ eV, whereas, in contrast, $|\epsilon_{\rm 3d\uparrow}|$
  increases in the same abrupt manner from  $|\epsilon_{3\rm d\uparrow}| \approx  8.4$ eV to $|\epsilon_{3\rm d\uparrow}| \approx 14$ eV. This is indicative
of sudden orbital swelling of the $\rm 4s$$\uparrow$ orbital into the outer well of the effective potential  but orbital contraction of the $\rm 3d$$\uparrow$ orbital, at  $r_{0} = 3.2$. Indeed, figure \ref{fig7} clearly
demonstrates said behaviour of the $\rm 4s$$\uparrow$ and $\rm 3d$$\uparrow$ orbitals.
The implication is that the outer well of the potential is not strong enough to accommodate the five $\rm 3d$$\uparrow$ electrons in the well. As the $\rm 4 s$$\uparrow$ electron moves into the outer well, at $r_{0} = 3.2$, nuclear screening for the $\rm 3d$$\uparrow$ orbital lessens, by virtue of which the attractive force of the nucleus on the orbital increases. This makes the $\rm 3d$$\uparrow$ orbital to contract. Interesting, however, is that the change in its
 energy is significant ($|\epsilon_{3\rm d\uparrow}| \approx  8.4$ eV at $r_{0}=4.8$ versus $|\epsilon_{3\rm d\uparrow}| \approx  14$ eV at $r_{0}=3.2$) whereas the spatial contraction is somewhat not.  At   smaller values of $r_{0}$, $r_{0} < 3.2$, the energy of the $\rm 4s$$\uparrow$ orbital is
practically independent of $r_{0}$. This is because, at $r_{0} \leq 3.2$, almost entire $\rm 4s$$\uparrow$ electron density is accommodated in the outer well of the potential, so that the confining potential practically stops
exerting any additional pressure on the $\rm 4s$$\uparrow$ orbital, thereby leaving it unchanged with decreasing $r_{0}$. In contrast, decreasing $r_{0}$ to values below $r_{0} =3.2$ starts somewhat noticeably
 pushing the $\rm 3d$$\uparrow$ level upward the inner well once again (moving it from $|\epsilon_{3\rm d\uparrow}| \approx 14$ eV to $|\epsilon_{3\rm d\uparrow}| \approx 12$ eV) but keeps compressing the orbital itself.

\section{Conclusion}

In this paper, a number of interesting effects in the behaviour of confined atoms  has been uncovered, illustrated and interpreted. Such effects as confinement induced orbital fusion an fission are characteristic features of specifically semifilled shell atoms. Indeed, these effects take place for spin-up and spin-down subshells ${n\ell}^{2\ell+1}$$\uparrow$ and ${n\ell}^{2\ell+1}$$\downarrow$ with the same values of $n\ell$ which originally
are split (both in the energy and radial coordinate) due to differences in exchange interactions of electrons from these subshells with only spin-up electrons from an unpaired semifilled subshell ${n'\ell'}^{2\ell'+1}$$\uparrow$.
When only the unpaired semifilled subshell ${n'\ell'}^{2\ell'+1}$$\uparrow$ gets swollen into the outer well of the effective potential, its radius starts exceeding by far the radii of other subshells of the atom. Therefore, exchange interaction of inner electrons with electrons from the swollen ${n'\ell'}^{2\ell'+1}$$\uparrow$ subshell reduces to a zero, to an excellent approximation. Correspondingly, ${n\ell}^{2\ell+1}$$\uparrow$ and ${n\ell}^{2\ell+1}$$\downarrow$ subshells
fuse together into a single ${n\ell}^{2(2\ell+1)}$ subshell. When, in addition to the $n'\ell'$$\uparrow$ orbital, both the $n\ell$$\uparrow$ and $n\ell$$\downarrow$ orbitals swell into the outer well of the effective potential as well,
the presence or absence of exchange interaction between $n'\ell'$$\uparrow$ and $n\ell$$\uparrow$ electrons, on the one hand, and $n'\ell'$$\uparrow$ and $n\ell$$\downarrow$ electrons, on the other hand, starts mattering again.
This results in fission of the ${n\ell}^{2(2\ell+1)}$ subshell into two separate subshells, ${n\ell}^{2\ell+1}$$\uparrow$ and ${n\ell}^{2\ell+1}$$\downarrow$, in the outer potential well. Clearly, the discussed effects are inherent to semifilled shell atoms. As for other effects - orbital breathing and re-ordering - the author does not see any principle obstacles for them to emerge in atoms with arbitrary filled subshells. In fact, orbital re-ordering was earlier predicted for
a number of strongly compressed atoms under an impenetrable confinement \cite{Compressed}.

In conclusion, possible objects, processes, problems and situations of ceratin relevance to the solved problem might be porous materials with atoms-intruders inside porous, or a problem of separation and diffusion of gas molecules through porous nano-carbon formations such, e.g., as graphene or nanotubes \cite{PCCP13} where intriguing quantum sieving effects take place \cite{PCCP13,Q_Sieving}, as well as an atom, or atoms of a material, under pressure of powerful
\textit{short} laser pulses where, due to predicted in this paper confinement induced effects, absorption, emission, or scattering of radiation might take a much different route than that expected on the basis of the knowledge on interaction of radiation with free atoms. More knowledgeable and inventive researchers, than the author, could find or suggest more ways for a practical application of said effects. Anyway, ratiocinating on this matter, the author's personal viewpoint on the primary value of the present work, at the given time instant,
can be best represented by the following passage borrowed from work \cite{JPC} (devoted to atoms confined inside impenetrable spheres): `\textit{... we regard the solutions as teaching us something about the subject, although such systems... (in the sense of Dirac's famous remark) pertain to philosophy rather than physics. ... we should not be too worried by this aspect...}'.

\ack
 This work was supported by NSF Grant no.\ PHY-$0969386$.

\newpage

\section*{References}
\begin{thebibliography}{}
%
\bibitem{Michels} Michels A, de Boer J and Bijl A 1937 \textit{Physica} \textbf{4} 981
%
\bibitem{Sommerfeld} Sommerfeld A and Welker H 1938 \textit{Ann. Phys.} {\bf 32} 56
%
\bibitem{JPC} Connerade J-P, Kengkan P and Semaoune R \textit{J. Chin. Chem. Soc.} 2001 \textbf{48} 265
%
\bibitem{RPC'04} Dolmatov V K, Connerade J-P, Baltenkov A S and Manson S T 2004 \textit{Radiat. Phys. Chem.} \textbf{70} 417
%
\bibitem{AQC57} Sabin J R, Br\"{a}ndas E and Cruz S A (ed) 2009 \textit{Advances in Quantum Chemistry: Theory of Confined Quantum
Sytems: Part 1}  vol \textbf{57} (New York: Academic) pp 1-334
%
\bibitem{AQC58} Sabin J R, Br\"{a}ndas E and Cruz S A (ed) 2009 \textit{Advances in Quantum Chemistry: Theory of Confined Quantum
Sytems: Part 2}  vol \textbf{58} (New York: Academic) pp 1-297
%
\bibitem{Dolmatov12} Dolmatov V K and King J L 2012 \jpb \textbf{45} 225003
%
\bibitem{Manson13} Haso\u{g}lu M F, Zhou H-L, Gorczyca T W and Manson S T 2013 \PR A \textbf{87} 013409
%
\bibitem{Cruz13} Cabrera-Trujillo R and Cruz S A 2013 \PR A \textbf{87} 012502
%
\bibitem{Auburn13} Lee T-G, Ludlow J A and Pindzola M S 2013 \PR A \textbf{87} 015401
%
\bibitem{G-Mayer} Goepper-Mayer M 1941 \PR \textbf{60} 184
%
\bibitem{Connerade_book} Connerade J-P 1998 \textit{Highly Excited Atoms} ( Cambridge: Cambrigde University Press)
%
\bibitem{Slater} Slater J C 1974 \textit{The Self-Consistent Field for Molecules and Solids} (New York: McGraw-Hill)
%
\bibitem{JETP83} Amusia M Ya, Dolmatov V K and Ivanov V K 1983 \textit{Zh. Eksp.
Teor. Fiz.} \textbf{85} 115 [1983 Sov. Phys. JETP \textbf{58} 67].
%
\bibitem{Cr93} Dolmatov V K 1993 \jpb \textbf{26} L393
%
\bibitem{Mn3s} Amusia M Ya and Dolmatov V K  1993 \jpb \textbf{26} 1425
%
\bibitem{Compressed} Connerade J-P, Dolmatov V K and Lakshmi P A 2000 \jpb \textbf{33} 251
%
\bibitem{PCCP13} Jiao Y, Du A, Hankel M and Smith S C 2013 \textit{Phys. Chem. Chem. Phys.} (accepted manuscript, published on http://pubs.rsc.org, doi: 10.1039/C3CP44414G)
%
\bibitem{Q_Sieving} Beenakker J J, Borman V D and Krylov S Y 1995 \textit{Chem. Phys. Lett.} \textbf{232} 379
%
\end{thebibliography}
\end{document}